\documentclass[10pt,twocolumn,twoside]{IEEEtran}
\usepackage[utf8]{inputenc}
\usepackage{hyperref}
\usepackage{graphicx}
\usepackage[table]{xcolor}
\usepackage{multicol}
\usepackage{subcaption,multirow,amsmath,bm}
\usepackage{float,verbatim}
\usepackage{subcaption,multirow}
\definecolor{Gray}{gray}{0.9}
\usepackage{booktabs} 
\usepackage{afterpage} 
\usepackage{stfloats}  

\definecolor{maroon}{cmyk}{0,0.87,0.68,0.32}
\title{Vocal Tract Length Warped Features for Spoken Keyword  Spotting} 
\date{June 2021}
\author{Achintya kr. Sarkar,  Priyanka Dwivedi, Zheng-Hua Tan, \emph{Senior Member, IEEE}
\thanks{A. K. Sarkar and P. Dwivedi are with the Indian Institute of Information Technology, Sri City, India (E-mail: sarkar.achintya@gmail.com). Z.-H. Tan is with the Department of Electronic Systems, Aalborg University, Denmark (E-mail: zt@es.aau.dk).}}

\begin{document}

\maketitle
\begin{abstract}
 In this paper, we propose several methods that incorporate vocal tract length (VTL) warped features for spoken keyword spotting (KWS). The first method, \emph{VTL-independent KWS}, involves training a single deep neural network (DNN) that utilizes VTL features with various warping factors. During training, a specific VTL feature is randomly selected per epoch, allowing the exploration of VTL variations. During testing, the VTL features with different warping factors of a test utterance are scored against the DNN and combined with equal weight.
 In the second method scores the conventional features of a test utterance (without VTL warping) against the DNN.
 The third method, \emph{VTL-concatenation KWS}, concatenates VTL warped features to form high-dimensional features for KWS.  Evaluations carried out on the English Google Command dataset demonstrate that the proposed methods improve the accuracy of KWS. 
 
\end{abstract}
\section{Introduction}
Spoken keyword spotting (KWS) is the task of identifying whether or not a word from a predefined list of words was uttered in a given speech signal \cite{lopez2021deep}. KWS has many potential applications such as audio indexing, home automation robotics, awake-up word detection in mobile devices, etc. Due to the natural variability in vocal tract length (VTL) among individuals, the spectral representation of the same word can differ when uttered by different speakers. This VTL variability affects the performance of automatic speech recognition systems, which aim to recognize the linguistic content of speech signals regardless of speaker variability. To mitigate the effect of VTL variability, vocal tract length warping was introduced in \cite{7078563,akhil-interspeech2008,Jaitly_vocaltract, Lee-Rose98} and  subsequently  applied in speaker recognition \cite{sarkar09_interspeech, 9339931}, and spoken term detection \cite{MADHAVI2019175}. The motivation behind this technique is to align the spectra produced by two speakers (A and B) for the same linguistic content. This alignment is achieved by scaling the frequency axis of the signal spectra as follows,
\begin{eqnarray}
S_A(f) = S_B(\alpha f) \label{eq:equation1}
\end{eqnarray}
where $\alpha$ is called the VTL warp/warping factor or VTL factor. 

To improve the accuracy of KWS, various deep neural networks (DNNs) have been investigated in the literature. These include the convolutional-recurrent neural-network-based multi-head attention mechanism \cite{deandrade2018neural, Rybakov_2020, Berg_2021}, vision transform (ViT) \cite{Berg_2021}, time delay neural network (TDNN) \cite{myer2018efficient} and emphasized channel attention, propagation and aggregation (ECAPA) TDNN \cite{lv2023dccrnkws}, temporal convolution ResNet (TC-ResNet) \cite{choi2019temporal},  broadcasted-residual (BC)-ResNet-based KWS \cite{kim21l_interspeech}, and feed-forward neural networks \cite{7178863}. Other variants of KWS systems and their applications are available in \cite{mrk2024noiserobustkeywordspottingselfsupervised, 10607847,yusuf2024pretrainingendtoendkeywordsearch,5700849}. Mostly KWS methods utilize Mel-frequency cepstral coefficients (MFCCs) as the 2D input to the network, often without considering the VTL warped features.

In this paper, we propose several KWS methods based on VTL warped features, all employing a DNN architecture. We investigate different approaches to utilize VTL features in KWS, resulting in three distinct methods:
\begin{itemize} 
\item[1)] \emph{VTL-independent KWS}: In this method, a single DNN is trained with all warped features sequentially, with a specific warped feature randomly selected per epoch. Since this method accumulates information from different VTL warped features into a single DNN framework (called VTL-DNN), we refer to this approach as \emph{VTL-independent KWS}. During testing, the different VTL warped features of a test utterance are evaluated against the VTL DNN, and the scores of different VTL features are combined in the score domain with equal weight.
\item[ 2)] \emph{VTL-independent$_{\alpha=1.00}$ KWS}: This method is analogous tothe \emph{VTL-independent KWS} with the only difference being that the conventional features of a test utterance (without VTL warping) are scored against the \emph{VTL-DNN}. 
 \item[3)] \emph{VTL-concatenation KWS}: In this method, the VTL warped features are concatenated resulting in a high-dimensional feature vector, which is used to train a DNN for KWS. During testing, the VTL concatenated feature of a test utterance is used to score against the trained DNN model. 
 \end{itemize}

We demonstrate the performance of the proposed methods on the English Google Command \cite{speechcommandsv2} dataset. The results demonstrate that the proposed methods achieve higher KWS accuracies compared to the conventional approach which does not utilize the VTL warped feature.

The paper is organized as follows. Section \ref{sec:proposedMethod} describes the proposed methods. The KWS classifiers are presented in Section \ref{sec:classifier}. The experimental setup is presented in Section \ref{sec:experiment_setup}. Section \ref{sec:result_discussion} analyzes the performance of the KWS. Finally, the paper is concluded in Section \ref{sec:conclusion}.

\section{Proposed Methods}
\label{sec:proposedMethod}
This section first introduces the concept of VTL warping factor and then describes the proposed methods.
\subsection{VTL warping factor}
The concept of VTL  warping factor $\alpha$ was first introduced in \cite{Lee-Rose98} to align the spectra of the same content uttered by two speakers by scaling the frequency axis of the signal spectra as in Eq. (\ref{eq:equation1}). 
Depending on the psychological structure of a person, the $\alpha$ value in Eq. (\ref{eq:equation1}) generally lies in between $[0.80, 1.20]$. In practice, a step size of $0.02$ is commonly used. This gives a set of $21$ $\alpha$ values. The  warped  spectrum $f_{wrp}^{(\alpha)}$ for a given signal having a spectrum $f$ is estimated piecewise linearly as 
\begin{eqnarray}
\small{
f_{wrp}^{(\alpha)}= 
\begin{cases}
    \alpha f,      &  0 \leq f \leq f_0\\
    \frac{f_{m} -\alpha f_0}{f_{m}-f_0}(f-f_0) + \alpha f_0,  & \quad f_0 \leq f \leq f_{m}
  \end{cases}
 }
\end{eqnarray}

\noindent where $f_{m}$ denotes the maximum bandwidth of the signal,and $f_0$ is an empirically chosen frequency.  

\noindent \subsection{VTL-independent KWS}
In this method, a single DNN is trained with all warped features. At each epoch, one warping factor is randomly selected, and warped features corresponding to the selected warping factor are used for training. A cross-entropy loss is used for training. A similar concept is utilized in \cite{Jaitly_vocaltract} for automatic speech recognition. 
During  testing,    all VTL warped features of a test utterance are scored against the single  VTL-DNN.  This yields a set of $21$ scores based on the number of VTL factor $\alpha$.  The scores of the test utterance for different VTL warped features are combined with equal importance. We call it \emph{VTL-independent KWS}.
 Equ. (\ref{eq: equation1}) depicts the score averaging across the VTL-independent KWS methods
\begin{eqnarray}
    FusionScore = \frac{1}{\#\alpha}\sum_{\alpha} \lambda_{vtl-dnn}(X_{\alpha}) \label{eq: equation1}
\end{eqnarray}
\noindent  $\lambda_{vtl-dnn}(X_\alpha)$ denotes the score of the VTL-DNN model $\lambda_{vtl-dnn}$ for the warped features  $X_\alpha$ of the test signal. The keyword class yielding the highest score is then identified as the recognized keyword.

\noindent \subsection{VTL-independent$_{\alpha=1.00}$ KWS}
This method is similar to the \emph{VTL-independent KWS}, with the key difference being that the unwarped VTL ($\alpha=1.00$)  features of the test utterance are scored against the VTL-DNN. 
It can be expressed as follows,
\begin{eqnarray}
    Score = \lambda_{vtl-dnn}(X_{\alpha=1.00}) \label{eq: equation2}
\end{eqnarray}
In terms of computational complexity and number of parameters, this method is analogous to conventional KWS. However, it can capture VTL feature information within a single DNN.

\noindent\subsection{VTL-concatenation  KWS}
In this method, all VTL warped feature vectors for a given speech signal are concatenated, resulting in high-dimensional feature vectors. For example, with $21$ values of $\alpha$ and each VTL $\alpha$ having a $40$-dimensional feature vector, this gives  $40\times 21$-dimensional feature vectors. The objective of this method is to investigate whether concatenated features are as effective or able to capture more relevant information compared to the \emph{VTL-independent} and \emph{VTL-independent$_{\alpha=1.00}$} KWS methods. 

\section{Classifiers}
\label{sec:classifier}
Four classifiers are considered for evaluating the performance of KWS methods. The first is GRU-MttAten \cite{deandrade2018neural, Rybakov_2020}, where  MFCC features  are processed through a 2D convolutional network, followed by a two-layer bidirectional long short-term memory (LSTM) network with a hidden layer dimension of $128$, incorporating gated recurrent unit (GRU) as in \cite{Rybakov_2020}  to jointly capture  temporal long-term dependencies. The feature representation in the middle of the bidirectional GRU layer is projected to a dense layer, serving as the query (Q) vector in the multi-head ($4$-head, i.e., $4$ Q matrices)  attention mechanism. Subsequently, the weighted average of the bidirectional GRU output is passed through three layers of a fully connected network before being projected at the output nodes.  The second classifier is the temporal convolution (TC)-ResNet8 architecture as per \cite{choi2019temporal}, consisting of three residual blocks and (16, 24, 32, 48) channels for each layer, including the first convolution layer. More details about these techniques can be found in \cite{Rybakov_2020,choi2019temporal}.

The third classifier employs the keyword transformer (KWT)-3 method, which leverages the concept of vision transformer (ViT) \cite{Berg_2021} in KWS In this approach, MFCC spectrograms are converted into patches to be fed into the transformer. 

The fourth one is based on BC-ResNet-8 based KWS \cite{kim21l_interspeech}, which leverages the advantages of both $1D$ and $2D$ convolutions, expanding the temporal output into the time-frequency domain.  The residual connection mapping enables the network to effectively represent useful audio information with less computation than convolutional networks. The core idea behind the broadcasted residual block is,
\begin{eqnarray}
y = x + BC(f1(avgpool(f2(x))))
\end{eqnarray} 
where $x$, $f_1$, $f_2$, and BC indicate the input feature, temporal and 2D operation, and broadcasting operation to expand the features into the original frequency dimension, respectively. The avgpool performs the average pooling in the frequency dimension. More details about the method can be found in \cite{kim21l_interspeech}.

\section{Experiment setup}
\label{sec:experiment_setup}
Experiments are conducted on the English Google command \cite{speechcommandsv2} dataset, which contains $35$ commands or keywords such as "up", "left", and "right". These $35$ commands are considered the target keywords, resulting in a total of $35$ classes for classification.  Table \ref{table:database} shows the number of speech files available for training and testing with each speech file being approximately $1$ second in duration. More details about the database can be found in \cite{speechcommandsv2}.

\begin{table}[h]
\caption{\it Number of class and examples in the training and evaluation sets.}
\begin{center}
\begin{tabular}{|l|c|c|c|}\cline{1-4} 
Database   & \# class     & Training   & Evaluation \\ \hline
English Google command     &  35          & 84843      & 11005    \\ \hline 
\end{tabular}
\end{center}
\label{table:database}
\end{table}

\begin{table*}[b]
\caption{\it Comparison of the proposed KWS methods against baseline methods on the Google Command dataset.}
\begin{center}
\begin{tabular}{|l|c|}\cline{1-2} 
Method              & Accuracy (\%))\\ \hline       
{\bf Baselines} (MFCC, $\alpha=1.0$):                           &       \\
TC-ResNet \cite{choi2019temporal}    & 94.02            \\
GRU-MttAten\cite{Berg_2021}         & 95.65        \\
KWT-3  \cite{Berg_2021}             & 96.20          \\
BCResNet-8 \cite{kim21l_interspeech} & 96.79         \\
                                                         &             \\
{\bf Proposed:}                                          &            \\
{\it Method 1:}                                          &            \\
VTL-independent-TC-ResNet                                      & {\bf94.61}      \\   
VTL-independent-GRU-MttAten                                    & {\bf96.22}         \\                                                   
VTL-independent-KWT-3                                          & {\bf96.51}         \\
VTL-independent-BCResNet-8                                     & {\bf 97.18}      \\ 
                                                         &           \\
{\it Method 2:}                                          &            \\
VTL-independent$_{\alpha=1.00}$-TC-ResNet                  & 94.29       \\
VTL-independent$_{\alpha=1.00}$-GRU-MttAten                & 96.07        \\   
VTL-independent$_{\alpha=1.00}$-KWT-3                      & 96.38        \\
VTL-independent$_{\alpha=1.00}$-BCResNet-8                 & 97.07        \\
                                                         &              \\
                                                        
{\it Method 3:}                                          &            \\
VTL-concatenation-KWT-3 & 96.10 \\

VTL-concatenation-BCResNet-8   &  96.05               \\  \hline 
                     
\end{tabular}
\end{center}
\label{table:kws}
\end{table*}

\begin{figure*}[b]
\hspace*{-0.2cm}\includegraphics[height=8.5cm,width=18.5cm]{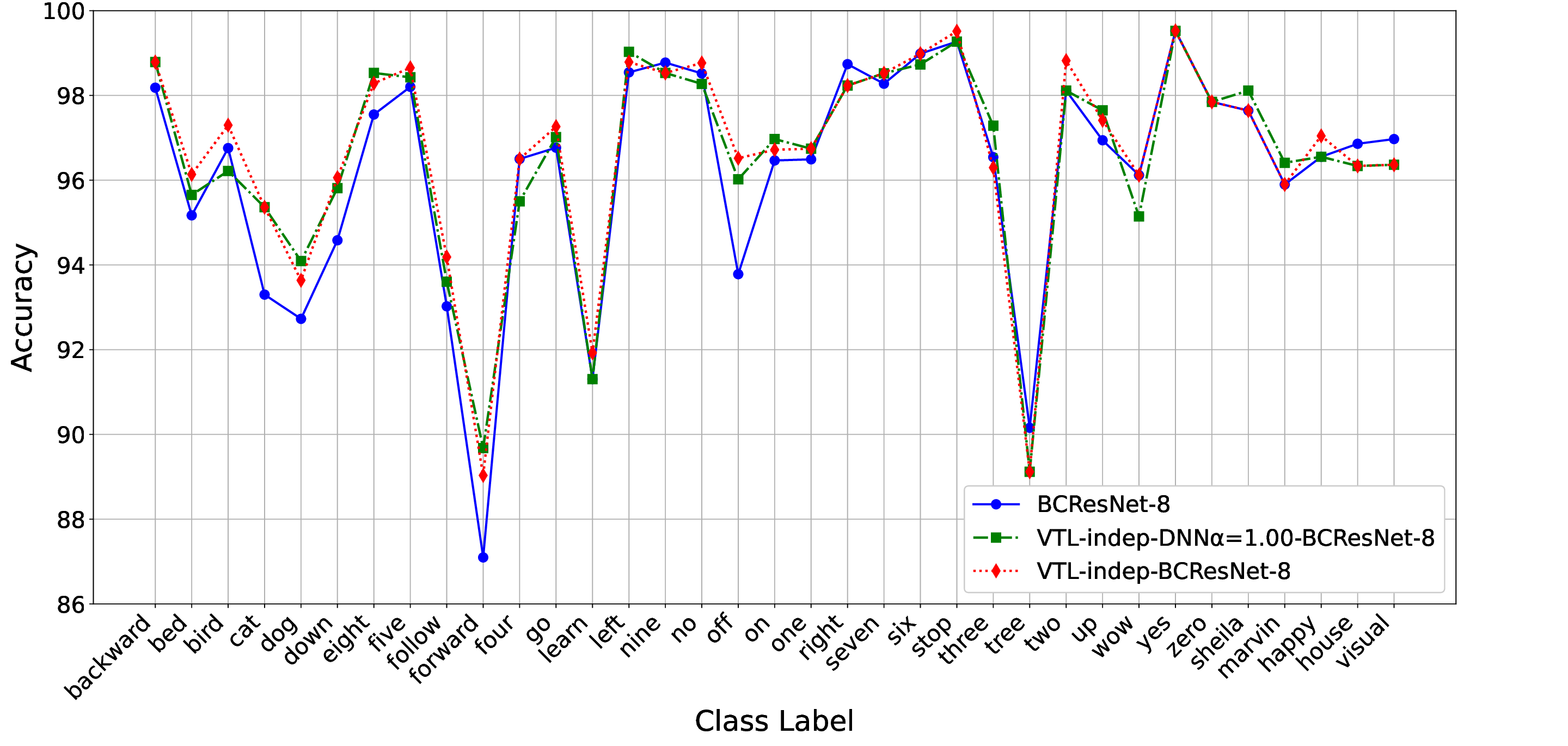}
\caption{\it Comparison of the class-wise performance of BCResNet-8,  VTL-independent$_{\alpha=1.00}$-BCResNet-8 and VTL-independent-BCResNet-8 based KWS methods on the Google Command dataset.}
\label{fig:class-wise-kws-Google}
\end{figure*}

During training the DNNs, data augmentation techniques are employed. These include time shift ([-100, 100] ms), resampling ([0.85, 1.15]), reducing background volume (0.1), applying time masks ([0, 25]), frequency mask ([0, 7]), and adding babble noise, reverberation, MUSAN music, and noise at $15$, $10$, $8$, $5$ dBs. The Kaldi \cite{Povey_ASRU2011} and librosa \cite{librosa} toolkits are used for data augmentation. 

The $40$ dimensional MFCC feature is extracted with a $30ms$ window at a $10ms$ frame shift using the HTK toolkit \cite{htkbook}. Regarding the config parameters, $f_0$ and  $f_{m}$ required for the VTL warped feature are set as $20$ Hz and $85\%$ of ($f$) Hz (the signal's maximum frequency), respectively.

All DNNs are trained from scratch for up to $100$ epochs with a weight decay value of $0.1$, a label smoothing value of $0.1$, the AdamW optimizer, and a batch size of $512$. The learning rate starts with $0.001$ and is updated using a cosine warmup schedule with a warmup value of $10$. 
For the VTL-DNN, the last epoch is trained using the feature ($\alpha =1.00$), which is the center value of all VTL factors, ensuring the method is unbiased towards the vtln warping factor ($\alpha \le 1.00$ and $\alpha \ge 1.00$) at the end. The Pytorch toolkit \cite{paszke2017automatic} is used to implement the DNNs. The seed values for the Numpy and Pytorch are set to $0$ and $42$, respectively, during model training to ensure reproducibility of the method's performance, unless specified otherwise for a particular method.

\section{Results and Discussions}
\label{sec:result_discussion}
In this section, we first compare the performance of the proposed KWS methods with the baselines on the English Google Command dataset. 
The results in Table \ref{table:kws} demonstrate that both the proposed \emph{VTL-independent} and \emph{VTL-independent$_{\alpha=1.00}$} methods, when combined with different DNN architectures, consistently outperform their baseline counterpart methods. This confirms the effectiveness of the proposed methods by incorporating VTL warped features in KWS.

Among these two proposed methods, \emph{VTL-independent} consistently proves to be superior. This indicates that utilizing VTL warped features in both training and testing are beneficial, as \emph{VTL-independent} uses these features in both processes while \emph{VTL-independent$_{\alpha=1.00}$} uses VTL warped features only during training. \emph{VTL-independent$_{\alpha=1.00}$}, on the other hand, has an advantage of maintaining the same computational complexity as the baselines during testing. 

The \emph{VTL-concatenation} method, however, does not perform as well as the baselines. This could be due to the significant increase in model size, which may result in inadequate training. 
Additionally, it is observed that among the baseline methods, \emph{BCResNet-8} performs the best. 

To gain further insights into the performance of these methods, we compare the class-wise performance of  \emph{BCResNet-8}, \emph{VTL-indeendent-BCResNet-8}, and \emph{VTL-independent$_{\alpha=1.00}$-BCResNet-8}. Figure \ref{fig:class-wise-kws-Google} shows that the proposed methods are superior in most classes.

\begin{figure}[ht]
    \centering
        \includegraphics[height=6.0cm,width=10.2cm]{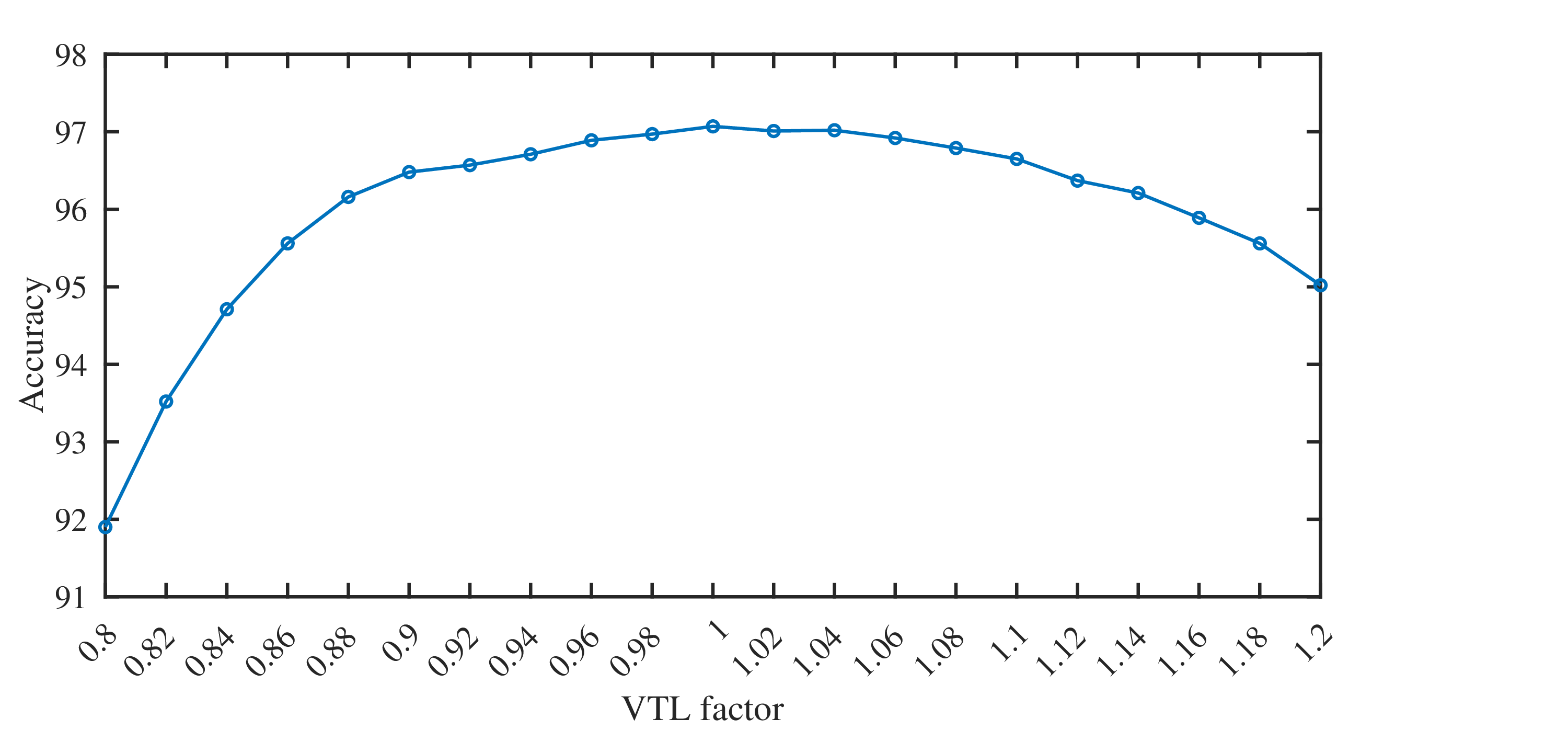} 
        \caption{\it Accuracy of KWS for different VTL warped factors during testing.}
    \label{fig:vtln-alpha-test-kws-Google}
\end{figure}

To investigate the impact of the different VTL warped features for the proposed KWS method during testing, we show the accuracies of KWS for the best-performing method, \emph{VTL-independent-BCResNet-8}, 
across VTL warping factors in  Fig. \ref{fig:vtln-alpha-test-kws-Google},  where we use only one single VTL factor at a time instead of using all of them. 
From Fig. \ref{fig:vtln-alpha-test-kws-Google},  it can be observed that the accuracy of the KWS initially increases as the value of VTL factor rises up to $\alpha=1.00$, after which it starts to decrease. This could be due to the random selection of the VTL-warped features during training for an epoch, leading to the feature closest to the mean of the VTL factors performing the best. 

\begin{table*}[!b]
\caption{\it Test accuracies (\%) and the $p$-values of Student's $t$-tests between the proposed KWS methods and the baseline method using the \emph{BCResNet-8} model on the  English Google Command dataset. The significance level was set as 0.05. The best results are marked in bold, and the statistically significant results are marked in italic.}
\begin{center}
\begin{tabular}{lll}
    \multicolumn{3}{c}{BCResNet-8}  \\ \hline
Baseline             &  VTL-independent                                 &   VTL-independent$_{\alpha=1.00}$        \\ 
Top-1 acc.           & Top-1 acc./$p$-value                               &  Top-1 acc./$p$-value       \\ \hline  \\
                    
96.82$\pm$ 0.05      &  {\bf \textit{97.04 $\pm$ 0.05/3.4718 $\times 10^{-5}$}} & \textit{96.96 $\pm$ 0.07/0.00162}     \\ \hline 
  
\end{tabular}
\end{center}
\label{table:kws_random}
\end{table*}


To test the statistical significance of the improvement brought by the proposed methods, we experiment with variants of the BCResNet-8 model, given their superior performance in their respective categories. We conduct experiments of initializing the model parameters with ten different random seeds based on  the computer's clock time.
Table \ref{table:kws_random} shows the mean accuracy $\pm$ $95\%$ confidence interval using the Student's $t$-test for the proposed methods and the baseline KWS method \emph{BCResNet-8}. 
The $p$-values for both proposed methods, \emph{VTL-independent} and \emph{VTL-independent$_{\alpha=1.00}$}, are less than  $0.05$, indicating the accuracy improvement obtained by the proposed methods is statistically significant.



\section{conclusion}
\label{sec:conclusion}

This paper proposed several methods that incorporate vocal tract length (VTL) warped features for spoken keyword spotting (KWS) and explored their combination with several deep neural network architectures.
We demonstrated that the proposed VTL-independent KWS method improves the KWS accuracy compared to conventional methods that do not consider warped features. The performance of the methods was validated using the English Google Command dataset. Future work includes exploring VTL-dependent features for personalized KWS.


\section{Acknowledgement}
A part of this work is supported by NLTM BHASHINI project funding 11(1)/2022-HCC(TDIL) from MeitY, Govt. of India.
        
\bibliography{references}
\end{document}